\documentclass[aps,pre,superscriptaddress,showkeys,floatfix]{revtex4} 
\usepackage{graphicx}
\usepackage{subfigure}
\usepackage{epsfig}

\newcommand{\bs}{\symbol{92}}

\begin{document}

\title{Optimal overlayer inspired by \textit{Photuris} firefly improves light-extraction efficiency of existing light-emitting diodes}

%\author{Annick Bay,$^{1,*}$ Nicolas Andr\'{e},$^{2}$ Micha\"{e}l Sarrazin,$^{1}$ Ali Belarouci,$^{3}$ Vincent Aimez,$^{3}$ Laurent A. Francis,$^{2,3}$ and Jean Pol Vigneron$^{1}$}

%\address{$^{1}$Research Center in Physics of Matter and Radiation (PMR), Department of Physics, University of Namur (FUNDP), 61 rue de Bruxelles, B-5000 Namur, Belgium \\$^{2}$ICTEAM Institute, Universit\'{e} Catholique de Louvain, B-1348 Louvain-la-Neuve, Belgium \\$^{3}$UMI3463- Laboratoire Nanotechnologies et Nanosystemes (UMI-LN2), Universit\'{e} de Sherbrooke – CNRS – INSA de Lyon – ECL – UJF-CPE Lyon, 2500 Boulevard de l'Universit\'{e}, Sherbrooke, Qu\'{e}bec J1K 2R1, Canada}

%\email{$^{*}$annick.bay@fundp.ac.be}

\author{Annick Bay}\affiliation{Research Center in Physics of Matter and Radiation (PMR),Department of Physics, University of Namur (FUNDP), 61 rue de Bruxelles, B-5000 Namur, Belgium}\email{annick.bay@fundp.ac.be}

\author{Nicolas Andr\'{e}}\affiliation{ICTEAM Institute, Universit\'{e} Catholique de Louvain, B-1348 Louvain-la-Neuve, Belgium}%\email{nicolas.andre@usherbrooke.ca}

\author{Micha\"{e}l Sarrazin}\affiliation{Research Center in Physics of Matter and Radiation (PMR),Department of Physics, University of Namur (FUNDP), 61 rue de Bruxelles, B-5000 Namur, Belgium}%\email{michael.sarrazin@fundp.ac.be}

\author{Ali Belarouci}\affiliation{UMI3463- Laboratoire Nanotechnologies et Nanosyst\`{e}mes (UMI-LN2), Universit\'{e} de Sherbrooke – CNRS – INSA de Lyon – ECL – UJF-CPE Lyon, 2500 Boulevard de l'Universit\'{e}, Sherbrooke, Qu\'{e}bec J1K 2R1, Canada}%\email{ali.belarouci@ec-lyon.fr}

\author{Vincent Aimez}\affiliation{UMI3463- Laboratoire Nanotechnologies et Nanosyst\`{e}mes (UMI-LN2), Universit\'{e} de Sherbrooke – CNRS – INSA de Lyon – ECL – UJF-CPE Lyon, 2500 Boulevard de l'Universit\'{e}, Sherbrooke, Qu\'{e}bec J1K 2R1, Canada}%\email{}

\author{Laurent A. Francis}\affiliation{ICTEAM Institute, Universit\'{e} Catholique de Louvain, B-1348 Louvain-la-Neuve, Belgium}\affiliation{UMI3463- Laboratoire Nanotechnologies et Nanosyst\`{e}mes (UMI-LN2), Universit\'{e} de Sherbrooke – CNRS – INSA de Lyon – ECL – UJF-CPE Lyon, 2500 Boulevard de l'Universit\'{e}, Sherbrooke, Qu\'{e}bec J1K 2R1, Canada}%\email{laurent.francis@uclouvain.be}

\author{Jean Pol Vigneron}\affiliation{Research Center in Physics of Matter and Radiation (PMR), Department of Physics, University of Namur (FUNDP), 61 rue de Bruxelles, B-5000 Namur, Belgium}%\email{jean-pol.vigneron@fundp.ac.be}

%--------------------------------------------------------------------------------------------------------------
%--------------------------------------------------------------------------------------------------------------

\begin{abstract}In this paper the design, fabrication and characterization of a bioinspired overlayer deposited on a GaN LED is described. The
purpose of this overlayer is to improve light extraction into air from the diode's high
refractive-index active material. The layer design is inspired by
the microstructure found in the firefly \textit{Photuris} sp. The actual dimensions and material composition have been optimized to take into account the high refractive index of the GaN diode stack. This two-dimensional pattern contrasts other designs by its unusual profile, its larger dimensions and the fact that it can be tailored to an existing diode design rather than requiring a complete redesign of the diode geometry. The gain of light extraction reaches values up to 55\% with respect to the reference unprocessed LED. 
\end{abstract}

%--------------------------------------------------------------------------------------------------------------
%--------------------------------------------------------------------------------------------------------------

%--------------------------------------------------------------------------------------------------------------
%--------------------------------------------------------------------------------------------------------------

\maketitle

\section{Introduction}

Optimization of light-emitting diode (LED) will be an important challenge for the
next few years. Today, much work focus on the choice of the material composition needed to
emit photons with a high efficiency \cite{review1,review2}. From this point of view, GaN-based LEDs present a
great interest. However, another independent way to improve the overall efficiency of LEDs
considers the light extraction after its emission at the p-n junction. Indeed,
transmission losses due to total internal reflection are responsible for low
light extraction efficiencies. Many surface structurations have been considered to overcome this specific
problem \cite{nano1,nano2,nano4,nano6,nano7,nano8,nano9,nano10,nano11,plasmon1}. For instance, roughness usually leads to diffraction, which provides new optical channels for light extraction. Nanostructures are also considered, to mimic a layer with
a graded refractive index \cite{gradient1,gradient2,gradient3}, another way to reduce abrupt refractive index contrast and increase the light extraction efficiencies. Finally, another approach considers microstructures which act as tiny lenses or
other optical devices to guide light out of the LED \cite{micro1,micro2,micro3,micro4,micro5,micro6}. In the present paper,
we suggest a new patterned device for the GaN-based LED. This pattern is directly
inspired by the microstructure found in the firefly \textit{Photuris} sp. (Lampyridae) \cite{OE_Photuris, SPIE_Photuris_09}, with the particularity to be created on an overlayer, added as a \textit{supplementary} layer, on an existing LED structure.

Fireflies produce their own light, which needs to be extracted from photocytes found in its bioluminescent organ (called \bs lantern''). It has been shown in previous works \cite{nano1,nano2,nano4,nano6,nano7,nano8,nano9,nano10,nano11,plasmon1,gradient1,gradient2,gradient3,micro1,micro2,micro3,micro4,micro5,micro6} that the firefly \textit{Photuris} sp. presents a two-dimensional pattern with an asymmetrical triangular profile with sizes in the range of a few micrometers on its lantern cuticle. This biological optical device increases the
extraction of the light chemically produced inside the insect's 	abdomen. Following a biomimetic approach, we transfer our understanding of the firefly structure into the LED context. 

In section \ref{theory}, we recall the interesting pattern found on the firefly and its influence on the light extraction efficiency. Next, the dimensions of the system are optimized according to
the optical properties of GaN, which are different from those in the biological tissues. Again the light extraction efficiency is precisely calculated. In section \ref{experiment}, an experimental device is described, fabricated and tested. The advantages of the present device over previous systems are also briefly discussed.

%--------------------------------------------------------------------------------------------------------------
%--------------------------------------------------------------------------------------------------------------

\section{Bioinspired light-emitting diode design}
\label{theory}

\subsection{The biological model}\label{subsec-II.1}

\begin{figure}[h]
\begin{center}
\epsfig{file=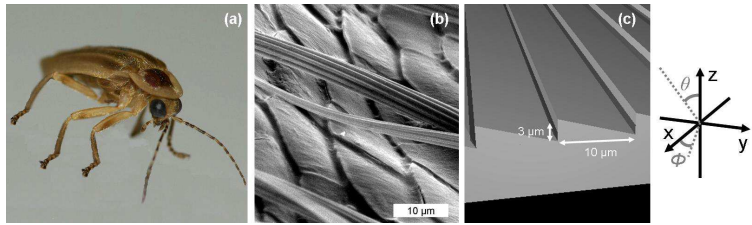,width=10cm} %height=4cm
\caption{(a) Firefly \textit{Photuris} sp. (b) SEM picture of the cuticle structuration above the bioluminescent organ. (c) Model used for simulations of the light propagation. Period = 10 $\mu m$. Height = 3 $\mu m$ }
\label{fig01}
\end{center}
\end{figure}

A former study on \textit{Photuris} has revealed an interesting structure on the outer cuticle of the firefly \cite{OE_Photuris,SPIE_Photuris_09}. Figure \ref{fig01}(b) shows a Scanning Electronic Microscope (SEM) image of the outer cuticle. The cuticle is made of scales bonded to create a continuous surface. The end of each scale is protruding in the direction of the abdominal tip. In the following, such a periodic structure is referred to as a factory-roof pattern. Different measurements of this tilted scale structure leads to average values of 10 $\mu$m for the period and 3 $\mu$m for the height of the protruding end. Figure \ref{fig01}(b) shows the model which was considered for the simulations of the light extraction \cite{SPIE_Photuris_09,SPIE_Photuris_12}. The model is a two-dimensional asymmetric model, which has two geometrical effects on the light extraction efficiency (LEE): (i) The tilted slope of the scale changes the orientation of the planar interface. (ii) A sharp edge is created at the end of the protruding scale.

\begin{figure}[h]
\begin{center}
\epsfig{file=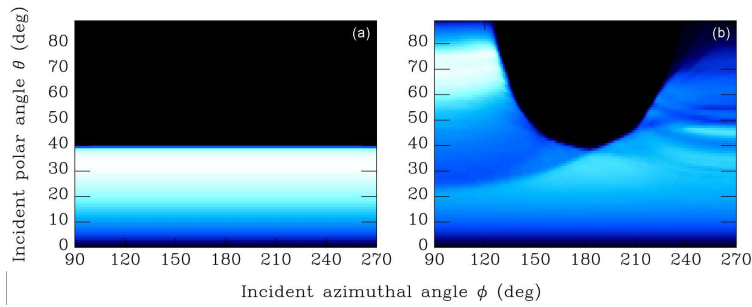,width=10cm} %height=5cm
\caption{LEE maps showing the integrated extracted light intensity as a function of the incident polar ($\theta$) and azimuthal ($\phi$) angles. Zero transmission is represented by black areas and maximal transmission is represented by white areas. (a) Plane surface. (b) Firefly structure.}
\label{fig02}
\end{center}
\end{figure}

The firefly is mainly made of chitin, which has an optical index of 1.56 \cite{chitin}. The emergent medium is air. Let us first consider the case of a planar interface. The critical angle $\theta_{crit}$ for total internal reflection has a value of 40$^{\circ}$. Past this angle, no light is transmitted from the chitin into air. The following simulations have been conducted with a home-made rigorous coupled-waves analysis (RCWA), using a transfer-matrix algorithm \cite{code01,code02}. Figure \ref{fig02}(a) shows the integrated light extraction efficiency (LEE) as a function of the incident polar ($\theta$) and azimuthal ($\phi$) angles. The polar angle $\theta$ is defined from the direction perpendicular to the interface, whereas the azimuthal angle $\phi$ is defined in the plane of this interface. One can easily see on Fig. \ref{fig02} that the transmission falls to 0\% at an incident polar angle of 40$^\circ$ (black area), which corresponds to the critical angle mentioned earlier. Due to this extraction confinement, only 20\% of the light can escape from the incident medium.

Considering now the specific structure of the scales as shown on Fig. \ref{fig01}(b), the light extraction dramatically changes. Figure \ref{fig02}(b) shows the result of the simulation with the factory-roof pattern. An azimuthal angle of $\phi=90^\circ$ corresponds to an incidence perpendicular to the sharp edges, $\phi=180^\circ$ corresponds to an incidence parallel to the protruding edges and $\phi=270^\circ$ describes an incidence perpendicular to the slope of the edges. The light extraction angular distribution is considerably changed with this corrugated surface. For $\phi=90^\circ$, the light extraction is not any more limited by the critical angle and the light can easily escape the incident medium above the critical angle. At an incident angle of $\phi=180^\circ$, incidence parallel to the protruding edges, the light extraction is again limited by the critical angle, because along this azimuthal incidence, the factory-roof pattern is invariant under all translations. For an azimuthal incidence $\phi=270^\circ$ the light extraction is again possible above the critical angle. The distribution is not exactly symmetric, as the scattered waves travel through the asymmetric profile along opposite directions. The LEE is now enhanced, up to 29\%. This specific structure extracts 45\% more light than a plane surface. Two effects are responsible for this large improvement: (i) The tilted scales can be considered as prisms that geometrically change the way the light rays impinge on the interface; (ii) Diffusion takes place at the sharp edges and creates a new channel for light transfer. This, as calculated, helps improve considerably the LEE.

%--------------------------------------------------------------------------------------------------------------

\subsection{LED bioinspired pattern and optimization}\label{subsec-II.2}

\begin{figure}[t]
\begin{center}
\epsfig{file=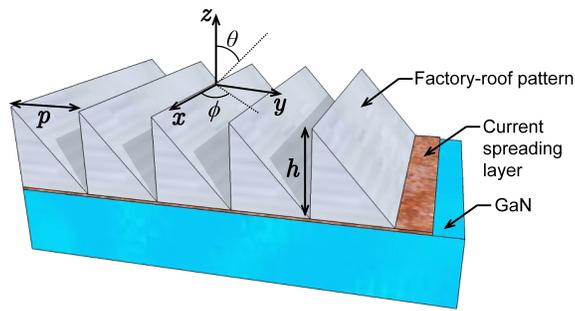,width=7.5cm}
\caption{Sketch of the device under theoretical consideration. The LED is
modelized by a GaN semi-infinite bulk endowed with a $10$ nm current
spreading layer made of Ni-Au alloy. The device is covered by a patterned
photoresist layer. The periodicity $p$ and the height $h$ of the factory-roofs
have to be optimized.}
\label{fig03}
\end{center}
\end{figure}

As explained in the previous section, the firefly \textit{Photuris} sp. presents an optimized structure regarding the light extraction. Therefore, one can expect learning from the firefly how to optimize the efficiency of some artificial light sources. Here, we consider a GaN-based LED covered by a photoresist which is patterned according to the morphology observed on the firefly (see Fig. \ref{fig03}).

\begin{figure}[h] 
\begin{center}
\includegraphics[width=10cm]{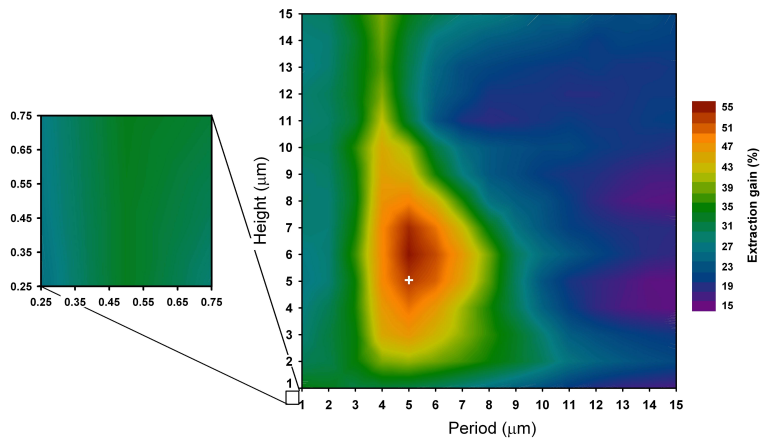}
\caption{Optimization of the photoresist pattern regarding the factory-roof period
and height. White cross corresponds to the setup under experimental consideration.}
\label{fig04}
\end{center}
\end{figure}

While the emission peak of the firefly bioluminescence corresponds to a wavelength
near $560$ nm, the main wavelength of the GaN LED is $425$ nm. The following calculations are performed at this specific emission wavelength. Of
course, the optical indexes of GaN and photoresist are different from those of the
chitin compounds in the firefly. As a consequence, the geometrical and the compositional parameters of the
artificial pattern will differ from those found in the insect. Then, considering the
relevant factory-roof pattern, we must compute the optimized parameters for
the LED. In the following, the GaN dielectric constant is taken from the analytical model of Ref. \cite{GaN} ($\epsilon=6.4$). The dielectric constant of the current spreading layer (made of nickel and gold alloy) at the emission wavelength is calculated from the metals dielectric constants found in Ref. \cite{Palik} ($\epsilon_{Ni}=-3.7+i8.1$, $\epsilon_{Au}=-1.6+i6.3$). The dielectric constant of the photoresist used in the following is
extracted from photoresist manufacturer's data ($\epsilon=2.763$ for photoresist $AZ$ $9245$\textregistered) \cite{Microchemicals}. Making use of a RCWA approach \cite{code01,code02}, we have performed numerical computations of the extraction gain for many
values of the period $p$ and of the height $h$ of the factory-roofs. The
light extraction is computed in each case as defined in the previous section.
The results are shown in Fig. \ref{fig04} where we present the
extraction gain, i.e. the relative deviation between the new device and a standard LED. The latter is simply a GaN bulk endowed with a spreading layer
but devoid of photoresist, and for which the absolute LEE
value is easily computed ($3.7$ \%). Referring to the firefly data and feasibility arguments, we prospect
for period and height varying between $1$ and $15$ $\mu $m.
The LEE clearly exhibits a peak at $p=5$ $\mu $m and $h=6$ $\mu $m. The gain then reaches 55\%, i.e. the patterned LED is theoretically 55\% more efficient than the classical
LED. Such an extraction enhancement at specific values of $h$ and $p$ is quite difficult to specify. Indeed, in the present device, geometrical optics and light diffusion prevail. This results into cooperative effects where each prism can couple to each others through diffracted rays of light or scattered light. This is a complex problem and it is difficult to obtain an intuitive or analytical interpretation which can justify for an optimal extraction at specific values of $h$ and $p$.

Such a result is really promising when considered against usual patterned devices
found in the literature addressing LED light extraction, where two kinds of devices are usually considered. The first one comprises subwavelength structures, which act as an optical adaption layer with an effective optical index gradient \cite{gradient1,gradient2,gradient3}. The other presents
micrometric patterns which do not exceed $1$ or $2$ microns. In those, LEE is improved through coherent multi-scattering processes \cite{micro5,micro6}. 
By contrast, the present large micrometric structure is novel in the landscape of light-extracting devices and is not intuitive. It is discovered thanks to the observation of the living world and proved through numerical simulations.

In the present work, the firefly-inspired structure is endowed with a very large pattern
which definitely contrasts previously suggested setups. Indeed, the pattern dimension under consideration ($p=5$ $\mu $m, $h=6$ $\mu $m) is almost twelve times larger than the wavelength of the LED light. In opposition, for values such that $1 \leq p,h \leq 2$ $\mu $m, LEE improvement is only about $30$ \%.
We also show the efficiency of the factory-roof pattern for nanometric scales usually encountered in previous works (see
enlargement map in Fig. \ref{fig04}). In this case, the greatest gain is about $33$ \% when $p\approx 500$ nm and $h\approx 500$ nm. As a consequence, with a LEE gain greater than $50$ \%, the firefly setup apperas to be widely more efficient than those conventional setups.
For fabrication convenience, a structure such that $p=5$ $\mu $m
and $h=5$ $\mu $m will actually be fabricated. In spite of this approximation, the gain is almost optimal with $%
54$ \% (see white cross in Fig. \ref{fig04}). In the next section, the processes of light extraction will be detailed, considering these parameters.

%--------------------------------------------------------------------------------------------------------------

\subsection{Theoretical light extraction}\label{subsec-II.3}

\begin{figure}[h]
\begin{center}
\epsfig{file=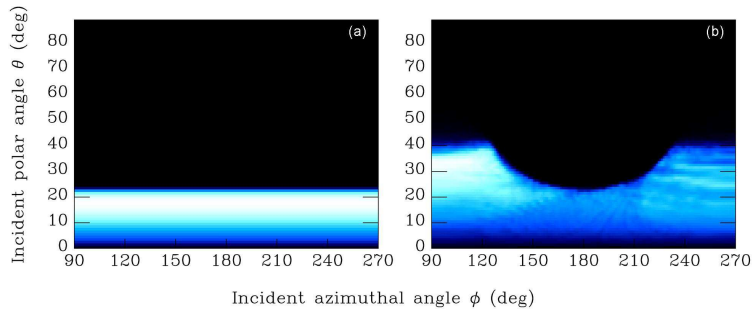,width=10cm} %height=5cm
\caption{LEE maps showing the integrated extracted light intensity as a function of the incident polar ($\theta$) and azimuthal ($\phi$) angles of the diode configuration. Zero transmission is represented by black areas and maximal transmission is represented by white areas. (a) Plane surface. (b) Firefly-like structure realized in photoresist.}\label{fig05}
\end{center}
\end{figure}

As seen in section \ref{subsec-II.2} the structure which is considered has a period of 5 $\mu$m and a height of 5 $\mu$m. The refractive index of the incident medium (GaN) is much higher than in the firefly case (2.53 and 1.56 respectively). The critical angle is therefore strongly reduced and light extraction falls to zero above 23$^\circ$. Figure \ref{fig05}(a) shows the integrated light extraction intensity as a function of the incident polar and azimuthal angle ($\theta$ and $\phi$ respectively). The light extraction is limited to incident polar angles below 23$^\circ$. The amount of extracted light is 3.7\%. Considering the case with the optimal patterned device made of photoresist on the LED, the light extraction behaviour changes again completely: in the azimuthal incident directions perpendicular to the sharp edge (i.e. $\phi=90^\circ$ and $\phi=270^\circ$), light can escape the incident medium above the critical angle. LEE is enhanced and reaches 5.7\%, so that 54\% is gained in comparison to our planar reference interface.

As a consequence we show that the use of a factory-roof pattern on an add-on photoresist overlayer is even more efficient for the artificial LED than for the firefly. In the following, we confirm these predictions by direct measurements. Note that the current spreading layer plays no relevant role in the process described here except for an overall attenuation of the extracted light. The light extraction would be better with a current spreading layer made with a less absorbing material than metal (ITO for instance \cite{Chong-PSS-2007}).

%--------------------------------------------------------------------------------------------------------------
%--------------------------------------------------------------------------------------------------------------

\section{Experimental part}
\label{experiment}

\subsection{LED microfabrication}
\begin{figure}[h]
\begin{center}
\includegraphics[height=4cm]{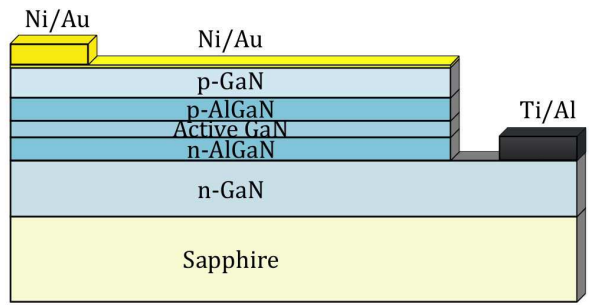}
\caption{AlGaN/GaN LED heterostructure cross-section.}
\label{fig06}
\end{center}
\end{figure}

In this work, the GaN-based emitting diode is the base on which bio-inspired photonic structures are added. The starting chip is composed of a sapphire substrate with a Hybrid Vapor Phase Epitaxy (HVPE-) grown AlGaN/GaN heterostructure as represented in Fig. \ref{fig06}. The specific layer thicknesses are listed in Table \ref{Table01}. The microfabrication processing requires four photolithographic steps to define etched and metalized areas. The first step uses a KMPR photoresist mask to form 1.5 $\mu$m mesas by a Cl$_2$/Ar inductively coupled plasma etching. After cleaning, a recrystallization annealing is then operated at 600$^\circ$C during 20 min to reduce the defects caused by the etching. The spreading layer, n-pad and p-pad are successively deposited by metal evaporations and lift-offs. The spreading layer deposition is required to increase the poor p-GaN electric performance. Several thermal annealings (1 min.) are performed after each lift-off. This procedure is important to ensure low-resistance ohmic contacts. The spreading layer annealing under air atmosphere, in place of N$_2$ atmosphere, oxidizes the Ni layer. These specific annealing conditions offer a compromise between the electrical conductivity and the optical transmittance, since the LED light has to pass through the spreading layer \cite{Ho}.  The annealing temperatures are respectively 580$^\circ$C, 550$^\circ$C and 500$^\circ$C.

\begin{table}[h]
\begin{center}
\caption{LED layers thicknesses.}
\begin{tabular}{|c|ccccc|}
\hline
 					& \textbf{p-GaN}  			& \textbf{p-AlGaN} 			& \textbf{active GaN} 		& \textbf{n-AlGaN} 				& \textbf{n-GaN}			\\
\hline
Thickness & 650 nm 			& 150 nm  			& 100 nm 				& 250 nm 					& 7$\mu$m		\\
\hline
					& \textbf{Sapphire} 		& \textbf{Ni/Au (p-pad)} & \textbf{Ti/Al (n-pad)} & \textbf{Ni/Au* (spread.)}& 					\\
\hline
Thickness & 430 $\mu$m 	& 20 nm/500 nm 	& 20 nm/500 nm 	& 5 nm/5 nm 			& \scriptsize{\textit{*before annealing}}				\\
\hline
\end{tabular}
\label{Table01}
\end{center}
\end{table}

%--------------------------------------------------------------------------------------------------------------

\subsection{Direct-writing laser patterning}

As discussed in Section \ref{subsec-II.1}, the micrometrical factory-roof shape produces a 54\% light extraction increase. With RCWA simulations, the optimal ratio between the period and height of the new bio-inspired structures is calculated (section \ref{subsec-II.2}). The optimal dimensions for this structure are changed in comparison to the firefly, but still remain micrometrical. This micrometrical dimension have a huge advantage in opposition to other nanometrical photonic structures found in nature. For instance, the reproduction of structures found on the wing of butterflies like \textit{Morpho rhetenor} or \textit{Papilio blumei} needs more complex techniques: multiple depositions, colloids etc. \cite{Vukusic,Kolle}. In the case of the firefly-microstructure fabrication, the use of standard optical lithography is well suited.

\begin{figure}[h]
\begin{center}
\includegraphics[height=3.5cm]{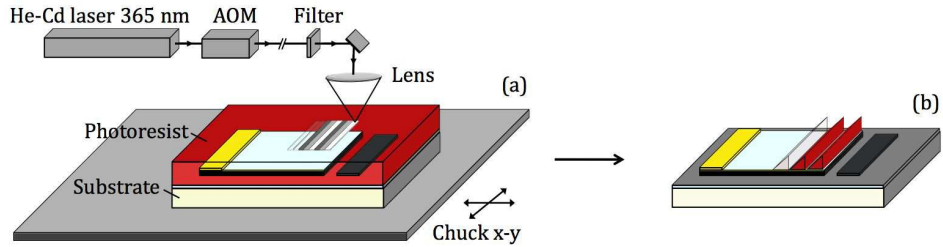}
\caption{Heidelberg photoplotter schematic for direct-writing laser strategy.}
\label{fig07}
\end{center}
\end{figure}

The factory-roof geometry of the abdominal firefly scales is built on photoresist by direct-writing laser lithography, as presented schematically in Fig. \ref{fig07}. In order to produce three-dimensional patterns in photoresist, the writing strategy of the DWL 66 Heidelberg photoplotter is divided in 5 main steps: (i) computer-aided layout design, (ii) photoresist spincoating and softbaking, (iii) sample alignment, (iv) power-modulated laser light exposure and (v) photoresist development. The He-Cd laser source emits at the wavelength of 365 nm and is modulated into a maximum of 32 power levels by an acousto-optic modulator (AOM). Optional filters are added to decrease the power level until the appropriate range is reached. The optical path ends with an interchangeable z-movable lens which focus the beam on the resist-coated sample. The 10 mm writing head unit is used in our case, allowing a field depth of 10 $\mu$m.

\begin{figure}[h]
\begin{center}
\includegraphics[height=4cm]{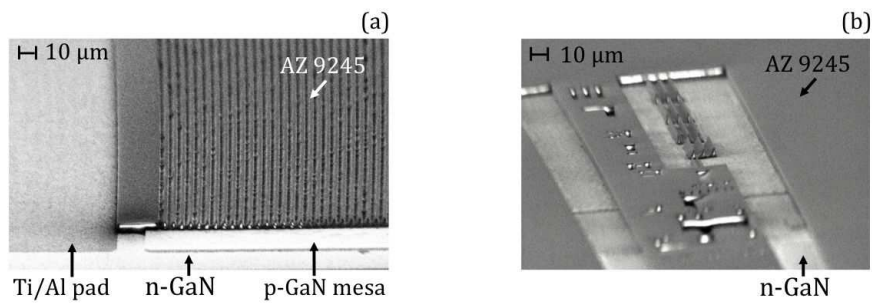}
\caption{Factory-roof patterns in $AZ$ $9245$\textregistered photoresist coating on a GaN-based LED, by direct-writing laser lithography.}
\label{fig08}
\end{center}
\end{figure}

The $AZ$ $9245$\textregistered photoresist allows the required 5 $\mu$m-thick coating with a small absorption coefficient, i.e. k = $2\times 10^{-4}$ and n = 1.65 at 435 nm after bleaching \cite{Microchemicals}. Working with a photoresist gives a fast and simple verification step to study the validity of the simulated designs on the LEE. Once the exposure dose and the concept are qualified, High Energy Beam Sensitive (HEBS) masks or resist micro-imprinting techniques can replace the time-consuming laser scanning by a one-time resist exposure or built-in. Figure \ref{fig08} presents two factory-roof microstructures on GaN light emitting diode materials.

%--------------------------------------------------------------------------------------------------------------

\subsection{Experimental light extraction}

A direct measurement of the light extraction efficiency has been
carried out, for comparison, on two identical diodes, with different
surface structures. For this purpose, two diodes were prepared on
the same support, with identical sizes and following the same
procedure. Both were coated with a uniform layer of photoresist, but only
one of them (diode A) was treated as explained above to produce a
factory-roof overlayer. The other one (diode B) is left with the
original flat photoresist surface to serve as a reference.

A scatterometer ELDIM EZContrast XL80MS was used to measure the
absolute radiance $R \left( \theta, \phi, \lambda \right)$ (in
watt/m$^2$/sr/nm) of both diodes at visible wavelengths $\lambda$,
in all directions above the diode surface, for polar angles $\theta$
ranging from 0 to 80$^\circ$ and azimuthal angles $\phi$ in the
range 0 to $360^\circ$. The diode was powered with a constant regulated current of 1mA. No specific polarization
state was selected. The power density per unit wavelength and unit
of emitting surface was calculated from these detailed data, by
integrating the radiance over a 2$\pi$ solid angle, as
\begin{equation}
P\left( \lambda  \right) = \int_0^{2\pi } {d\phi } \int_0^{{\theta
_{\max }}} {\sin \theta d\theta } \;R\left( {\theta ,\phi ,\lambda }
\right)
\end{equation}

The wavelength dependence of the radiance is preserved in the
integrated power, reflecting the spectral profile of the
electroluminescent emission. More precisely, 31 narrow (13
nm) bandpass filters were placed, in turn, on the path to the CCD
two-dimensional photon sensor. The exact transmission of these
filters is accounted for in the determination of the absolute recorded power value. The radiance delivered by both diodes was
measured independently. 
%Each of the diodes is constituted as a stack of semiconducting layers and both are covered by a 5~$\mu$m-thick layer of photoresist.

\begin{figure}[t]
\begin{center}
\includegraphics[width=7.5cm]{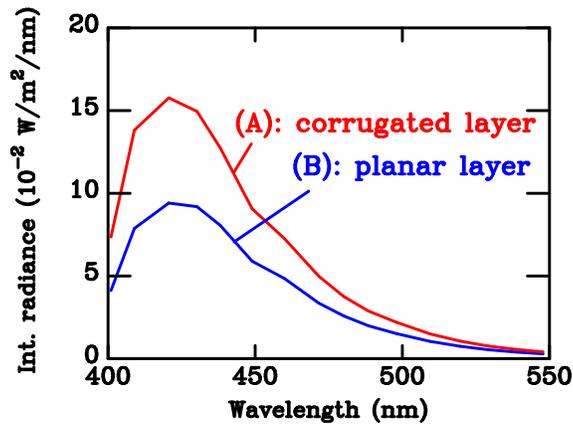}
\caption{Measured radiance integrated over the full 2$\pi$
hemispheric solid angle for two identical diodes covered by a photoresist
layer terminated (A) by a micrometric factory-roof corrugation
and (B) by a flat surface. The corrugation causes an 68\% increase
of the emitted power.} \label{fig09}
\end{center}
\end{figure}

The result of the measurement is presented in Fig. \ref{fig09}, where the
integrated radiance is shown as a function of the wavelength. The
emission peak is, for both (A) and (B) diodes, located at a wavelength close to
425 nm. The maximum emission power reaches 15.8$\times 10^{-2}$
W/m$^{2}$/nm for the corrugated surface and 9.4$\times 10^{-2}$
W/m$^{2}$/nm for the flat, reference, surface. This corresponds to a
68\% power increase, entirely due to the corrugating treatment of
the photoresist surface. As explained above, the improvement is essentially
caused by the diffusion at the sharp edges of the factory-roof
corrugation of the photoresist covering the diode. This 68\% value can not be compared immediately to the calculations of section \ref{subsec-II.3}, where a bare LED was considered, without any coated photoresist. For the experiment a photoresist coated patterned LED was compared with a photoresist coated unpatterned LED (photoresist-thickness of 5$\mu$m each). As these values doesn't correspond exactly to the former theoretical values, new simulations have been done to match the experiment. When comparing the extraction intensity of an LED simply coated with 5 $\mu m$ of photoresist to an LED coated with 5 $\mu m$ of photoresist and patterned with the optimal structure, the extraction gain reaches 33\%. This value is smaller than the 54\% calculated earlier (section \ref{subsec-II.3}, comparision to the bare LED). In fact, the photoresist layer gives already a slight improvement due to a gradual adaption of the refractive index: this system is 16\% more efficient than the bare LED (these values are summarized in Table \ref{Table02}). 

\begin{table}[h]
\begin{center}
\caption{Summary of extraction gains. A \bs bare LED'' is defined as a non-treated LED. A \bs coated LED'' is defined as a \bs bare LED'' with a 5 $\mu m$ thick photoresist layer. A \bs coated patterned LED'' is defined as a \bs coated LED'' where the photoresist is patterned with the specific factory-roof structure.}
\begin{tabular}{|c|c|}
\hline
coated patterned LED vs. bare LED & + 54\% \\
\hline
coated patterned LED vs. coated LED & + 33\% \\
\hline
coated LED vs. bare LED & + 16\% \\
\hline
\end{tabular}
\label{Table02}
\end{center}
\end{table}

Two aspects have to be considered to understand this acceptable difference between the theory and the measurements: (i) The patterning of the photoresist by direct laser-writing could have caused more, non-predictable structurations of the surface which improve even more the LEE than shown by the calculations. There remains a little, acceptable discrepancy between the actual device and the model. (ii) Even if the LEDs have been created under the exact same conditions, on the same support, there might be a little difference in the actual intensity of these two diodes. Regarding those considerations, we can only expect to have an even better LEE gain for the measurements with a bare LED.
These first measurements indicate that the tendencies predicted by the simulations are actually supported by a direct measurement.

\section{Conclusion}

Besides the internal quantum yield, the light extraction efficiency
is an important factor of all solid-state illumination sources. An
increased value of this factor has also been recognized as an
evolutionary advantage for some species of fireflies, with the
consequence that a structural improvement of the geometry of the
lantern surface was selected and well-established in modern insects.

The observed structure was used as an inspiration for the design of
a corrugated coating intended to improve the efficiency of a
standard GaN diode \textit{after} its fabrication. The proposed
system is a corrugated layer of photoresist deposited on the diode. The corrugation profile is directly inspired from the
geometry of the firefly surface cuticle, here described as a factory-roof two-dimensional profile. One new aspect of this
corrugation is its large size : 5~$\mu$m for the lateral period and
5~$\mu$m for the vertical height. This contrasts with other optimization
routes, where a submicron scale is ordinarily chosen. With this factory-roof profile, detailed computer simulations have shown that the
optimal length scale requires a few microns, both for the lateral
and vertical dimensions.

The fabrication technique applied in order to obtain corrugated overlayer has the advantages of simplicity and
scalability: a homogeneous layer of photosensitive resist is first
deposited by spin coating on the whole surface of the wafer and left
to harden; then, on each diode, appropriate radiation penetrates the
photoresist to a depth proportional to intensity, the illuminated region
being then easily removed. This technique allows for the production
of any profile on a precise area of a large surface.

The direct measurement of the emitted power showed an increase of
68\% with respect to the corresponding uncorrugated surface. Simulated light extraction led to a gain of 55\% in comparison to a plane surface. Those results are even more effective than in the natural system (\textit{Photuris} sp. firefly) which is the actual source of inspiration for this work.
Such an increase of efficiency is welcome: to produce photons with
less consumption of energy is one of the societal challenges that
should be met in the very next years.

\section*{Acknowledgments}
The authors acknowledge the financial support from Nanoquebec for this project, and thank the Sherbrooke cleanroom staff for their support. A. B. was supported as PhD student by the Belgian Fund for Industrial and Agricultural Research (FRIA). The project was partly funded by the \bs Action de
Recherche Concert\'{e}e'' (ARC) Grant No. 10/15-033 from the French
Community of Belgium. The authors also acknowledge using resources
from the Interuniversity Scientific Computing Facility located at
the University of Namur, Belgium, which is supported by the
F.R.S.-FNRS under convention No. 2.4617.07.

%--------------------------------------------------------------------------------------------------------------
%--------------------------------------------------------------------------------------------------------------

\end{document}